 \documentclass[final,5p,times,twocolumn,authoryear]{elsarticle}

\usepackage{amssymb}
\usepackage{lipsum}
\usepackage[dvipsnames]{xcolor}
\usepackage{colortbl}
\usepackage[round,authoryear]{natbib}   % or 'authoryear' if you want that style
\usepackage[colorlinks=true,
            citecolor=blue,
            linkcolor=blue,
            urlcolor=blue]{hyperref}

\usepackage{graphicx} % for figures
\usepackage{multirow} % for multirow cells
\usepackage{array}    % for better column formatting
\usepackage{booktabs} % for nice table lines
\usepackage{amsmath}
\usepackage{ulem}
\usepackage{comment}
\newcommand{\gPLUTO}{{\fontfamily{cmtt}\selectfont gPLUTO}}

\journal{Astronomy $\&$ Computing}

\begin{document}

\begin{frontmatter}



% Original:
%\title{GPU-enabled magneto-hydrodynamics for complex radio galaxy evolution}

% Version 1:
%\title{GPU-enabled Numerical Simulations of Radio Galaxy Evolution with the gPLUTO Code}

% Version 2:
% \title{Preliminary GPU-Enabled Magneto-Hydrodynamic Simulations of Radio Galaxy Evolution with gPLUTO}

% Version 3:
%\title{Application of the gPLUTO code for GPU-enabled relativistic MHD modeling of  Radio Galaxy Evolution with the gPLUTO Code}

% Version 4:
%\title{Exploring Radio Galaxy Evolution with the gPLUTO code: First Results from a GPU-Enabled MHD Framework}

% Version 5:
\title{Performance assessment of the \gPLUTO\, code for the numerical modeling of radio galaxy evolution}

\author[first]{Gourab Giri}
\affiliation[first]{organization={Istituto Nazionale di Astrofisica (INAF) – Istituto di Radioastronomia (IRA)},%Department and Organization
            addressline={via Gobetti 101}, 
            city={Bologna},
            postcode={40129}, 
            country={Italy}}
\author[second]{Andrea Mignone}
\author[second,inaf,icsc]{Alessio Suriano}
\author[second]{Marco Rossazza}
\author[second]{Stefano Truzzi}
\affiliation[second]{organization={Dipartimento di Fisica, Università degli Studi di Torino},%Department and Organization
            addressline={Via Pietro Giuria 1}, 
            city={Torino},
            postcode={10125}, 
            country={Italy}}
\affiliation[inaf]{organization={INAF - Istituto Nazionale di Astrofisica, Osservatorio Astrofisico di Torino},%Department and Organization
            addressline={Strada Osservatorio, 20}, 
            city={Pino Torinese},
            postcode={10025},
            country={Italy}}
\affiliation[icsc]{organization={ICSC - Italian Research Center on High Performance Computing, Big Data and Quantum Computing},%Department and Organization
            addressline={via Magnanelli, 2}, 
            city={Casalecchio di Reno},
            postcode={40033},
            country={Italy}}

\begin{abstract}
High-resolution tri-axial simulations are indispensable for realistically co-modeling the dynamical signatures and the radiative fingerprints of astrophysical jets, which are becoming increasingly important in modern computational studies of jet physics. 
However, such simulations impose extreme computational requirements that often exceed the capabilities of conventional CPU-based codes. 
GPU-accelerated simulations offer a transformative solution to mitigate these limitations. In this work, we present a detailed performance benchmarking of the recently developed GPU-enabled PLUTO code (\gPLUTO), demonstrating runtime speed-ups ranging from an order of magnitude to (approximately) over 30 relative to CPU-only configurations. 
A direct comparison between computations of extragalactic jet propagation performed at different grid resolutions confirm the physical fidelity and production readiness of the \gPLUTO\, code, while underscoring the importance of resolving the jet radius adequately to capture the jet dynamics accurately. Leveraging GPU-PLUTO’s capabilities, we finally present an application by performing high-resolution simulations of giant radio galaxy jets (GRGs $\gtrsim 1$ Mpc), representing the first such well-resolved 3D study to our knowledge (resolving scales down to 500 pc). 
These simulations probe a range of environmental effects on GRG jets, clarifying their formation from central galaxies within host cosmic structures, rapid peripheral expansion, and the development of asymmetric cocoon morphologies.
\end{abstract}




\end{frontmatter}

%\tableofcontents



\section{Introduction}
\label{Sec:Introduction}

Following the early exploration of extragalactic jets via synchrotron radiation in the 1950s \citep[e.g.,][]{Jennison1953}, the field underwent a transformative expansion in the 1970s, as reflected in a new wave of discoveries and systematic observations. 
This period marked not only the identification of multiple distinct classes of radio jets but also the development of statistical catalogs (e.g., X-shaped radio galaxies \citep{Hogbom1979}, giant radio galaxies \citep{Willis1974}, and other bent or peculiar sources \citep{Ekers1978}). 
The growing diversity in observed jet propagation mechanisms highlighted the need for theoretical frameworks capable of testing and extending the standard jet paradigm \citep[cf.][]{Hardcastle2020}. 
Early efforts focused on simplified analytical models, often restricted to one-dimensional treatments \citep[e.g.,][]{Hjellming1981,Gower1982}. 
However, assumptions such as constant parameter evolution (e.g., assuming a steady jet-head advance speed across diverse propagation scales) and the inability to reproduce the observed asymmetric and complex morphologies (jet regimes shaped by jet–environment coupling), soon revealed the limitations of such approaches. 
These shortcomings drove the field toward numerical simulations, enabling a more realistic exploration of the physical processes governing radio galaxy evolution \citep[cf.][]{Ferrari1998}.

The early phase of numerical modeling in this field began in the late 1980s, focusing primarily on 2D simulations of astrophysical jets \citep{Norman1982,Stone1992,Stone1993}. 
However, it was soon recognized that 2D setups could not fully capture the complex evolution inherent to jets \citep{Stone1994,Norman1996}, leading to significant differences compared to full 3D simulations \citep[see also,][]{Mignone2010}. 
This realization established a crucial numerical benchmark (requirements of full three-dimensional simulations) that has guided subsequent simulation efforts ever since \citep{Hardee2001,Nakamura2001}. 
While newly developed numerical techniques are often initially tested in 2D for computational efficiency \citep{Mignone2006,Vaidya2018,Kundu2022}, accurately capturing the dynamics of realistic cosmic jets demands full 3D implementation.

As numerical simulations of extragalactic jet propagation continue to advance \citep[cf.][]{Marti2019}, the importance of achieving high spatial resolution has become increasingly evident, especially to resolve the jet injection region with an adequate number of grid cells. 
In fact, high resolution not only improves the fidelity of jet propagation dynamics \citep{Costa2025} but also enables the accurate treatment of magneto-hydrodynamical (MHD) instabilities \citep{Wang2023}, entrainment processes \citep{Rossi2024}, and the influence of ambient medium tri-axiality on jet evolution \citep{Hodges-kluck2011}. 
Past studies have indicated that resolving at least eight to ten grid cells across the jet radius is necessary to capture the key dynamical features of jet propagation \citep{Stone1994,Norman1996,Aloy1999}. 
However, more modern works \citep{Mignone2010,Hodges-kluck2011,Giri2022_Sshape} demonstrated that increasing the resolution further to have $\sim 15$ grid cells within the jet radius substantially improves the fidelity of the simulations, enabling a more accurate representation of transverse structures (to the flow), shock formation along the jet beam, and pressure-driven morphological evolution.

In the current era, numerical simulations are becoming increasingly capable of modelling both jet propagation dynamics and their associated non-thermal radiative signatures in a self-consistent manner. 
This progress has been enabled by advanced Lagrangian–Eulerian particle–fluid techniques \citep[e.g.,][]{Mignone2018,Ogrodnik2021}, which allow the production of synthetic emission maps that can be directly compared with contemporary observational data \citep[e.g.,][]{Vaidya2018,Yates-jones2022}.

By allowing macro-particles (an ensemble of non-thermal particle distribution following a power-law) to track processes such as particle re-acceleration and cooling, these methods offer a powerful framework to link the jet dynamics with the observable non-thermal emission. 
Such approaches are in growing demand, not only to address modern questions surrounding radio galaxies \citep[e.g., see,][]{Giri2022_Xshape,Shabala2024,Dubey2024}, but also to refine our broader understanding of jet microphysics and the mechanisms governing their radiative fingerprints \citep[cf.][]{Peer2014}. 

Incorporating this suite of physical processes is both timely and essential for capturing the full complexity of jet–environment interactions. 
However, doing so drives simulations into a regime of unprecedented computational demand, rapidly exceeding the capabilities of traditional CPU-based codes. 
A recent study by \citet{Dubey2023} 
%{\color{red}[AM: MOVE REFERENCE HERE]} 
has shown that reliable emission modeling requires a resolution of roughly 25 grid cells per jet radius—a level of detail that, while feasible for compact or galactic-scale jets (tens of kpc), becomes increasingly prohibitive for classical doubles \citep[hundreds of kpc; cf.][]{Saikia2022} and giant radio galaxies \citep[>700 kpc; cf.][]{Dabhade2023}. 
As a result, many CPU-based studies have been forced to compromise: either truncating their models to the early phases of jet evolution \citep{Mukherjee2021} or reducing the physical realism by omitting key microphysical processes \citep{Giri2025}. 
This growing resolution–physics bottleneck highlights an urgent need for next-generation numerical approaches capable of handling both scale and complexity without compromise.

GPU-accelerated codes represent a crucial step in this transition, offering the ability to sustain both high spatial resolution and rich physical complexity over unprecedented spatiotemporal domains \citep[see also,][]{Stone2024,Liska2022}. 
By exploiting massively parallel architectures, these frameworks deliver significant speedups over conventional CPU-based approaches, enabling detailed simulations of multi-Mpc jets \citep[e.g., the observed systems of ][]{Andernach2025} without compromising physical realism. 
Codes such as \gPLUTO \, \citep[the GPU-accelerated successor to the widely used PLUTO legacy MHD code;][]{Mignone2007} exemplify this new generation of numerical tools—poised to bridge the gap between small-scale, well-resolved jet models and realistic large-scale evolutionary scenarios, which will be discussed in this work.

In Section~\ref{Sec:Governing equations and numerical implementation}, we provide a brief overview of \gPLUTO \, and outline its implementation for simulating systems ranging from classical doubles to giant radio galaxies. 
Section~\ref{Sec:Results} begins with a performance benchmarking analysis (\S~\ref{Subsec:Performance Benchmarking of (CPU vs GPU) PLUTO}), comparing \gPLUTO \, (on GPUs) with its CPU counterpart, followed by an investigation of jet dynamics across varying resolutions (\S~\ref{Subsec:Convergence analysis through resolution studies}). 
We then present a scientific application to giant radio galaxies—a domain where high-resolution 3D simulations remain exceedingly rare (\S~\ref{Subsec:Modeling Giant Radio Galaxies with GPU-Boosted Simulations}). Finally, in Section~\ref{Sec:Summary}, we summarize the key results and implications of this study. 

%%%%%%%%%%%%%%%%%%%%%%%%%%%%%%%%%%%%%%%%%%%%%%%%%%%%%%%%%%%%%%
\section{Governing equations and numerical implementations}
\label{Sec:Governing equations and numerical implementation}
%%%%%%%%%%%%%%%%%%%%%%%%%%%%%%%%%%%%%%%%%%%%%%%%%%%%%%%%%%%%%%

We first provide a concise overview of the physical model in the \gPLUTO\, code, outlining its core principles and governing equations. 
We then describe the setup of our astrophysical jet simulations and establish the primary scientific objectives of this work. 
The present paper is not intended as a code-description paper, but rather as a science application and science-readiness demonstration of \gPLUTO\, in the context of large-scale relativistic jet propagation.
A detailed description of the GPU-oriented implementation, code architecture, and performance scaling of the fluid module is provided in \citet{Rossazza2025}, which constitutes the foundational gPLUTO paper. The development and validation of the particle framework are described in Suriano et al. (\textit{submitted}).

\subsection{Equation models in the \gPLUTO}
%%%%%%%%%%%%%%%%%%%%%%%%%%%%%%%%%%%%%%%%%%%%%%%
The density of a relativistic flow in the observer frame is defined as $D = \gamma \rho$, where $\rho$ is the comoving rest-mass density and $\gamma$ is the Lorentz factor corresponding to the bulk velocity of the flow. 
We adopt a Cartesian coordinate system for our simulations, in which the three-velocity vector is denoted by $\mathbf{v}$. 
The covariant magnetic field four-vector is expressed as $b^\mu = [b^0, \mathbf{b}] = [\gamma \mathbf{v} \cdot \mathbf{B},\, \mathbf{B}/\gamma + \gamma (\mathbf{v} \cdot \mathbf{B})\mathbf{v}]$, where $\mathbf{B}$ is the magnetic field measured in the observer’s frame. 
The total (gas plus magnetic) pressure of the system is given by $P_t = P_g + \mathbf{B}^2/2\gamma^2 + (\mathbf{v} \cdot \mathbf{B})^2/2$, where $P_g$ denotes the gas pressure. 
Using these definitions, the total enthalpy can be written as $w_t = \rho h + \mathbf{B}^2/\gamma^2 + (\mathbf{v} \cdot \mathbf{B})^2$, with $h$ representing the specific enthalpy of the relativistic gas. 
The momentum density is then defined as $\mathbf{m} = w_t \gamma^2 \mathbf{v} - b^0 \mathbf{b}$, and the total energy density as 
$E_t = w_t \gamma^2 - (b^0)^2 - P_t$. 
We further define $I$ as the identity matrix. 

%With these definitions, the conservative form of the {\color{PineGreen} ideal} relativistic magnetohydrodynamic (RMHD) equations can be expressed as \citep{Komissarov1999},
%
%\begin{equation}
%  \frac{\partial \mathbf{U}}{\partial t} + \nabla \cdot \mathbf{T}(\mathbf{U}) = 0,
%\end{equation}
%
%where $\mathbf{U} = [D,\, \mathbf{m},\, E_t,\, \mathbf{B}]$ represents the vector of conserved quantities, and $\mathbf{T}(\mathbf{U})$ is the corresponding flux tensor. 
%{\color{red}[AM: YOU'RE REPEATING THE EQUATION TWICE... WE CAN SHORTEN THIS]}
With these definitions, the governing equations for a relativistic magnetized gas, as solved in our simulations, are therefore formulated under the assumption of an ideal RMHD framework employing the Taub–Mathews equation of state \citep{Taub1948,Mignone2007}, as,

\begin{equation}
\frac{\partial}{\partial t}
    \begin{pmatrix}
    D\\\textbf{\textit{m}}\\E_t\\\textbf{\textit{B}}
    \end{pmatrix}
    + \nabla \cdot
    \begin{pmatrix}
    D\textbf{\textit{v}}\\
    w_t\gamma^2\textbf{\textit{v}}\textbf{\textit{v}}-\textbf{\textit{b}}\textbf{\textit{b}}+IP_t\\
    \textbf{\textit{m}}\\
    \textbf{\textit{vB}}-\textbf{\textit{Bv}}
    \end{pmatrix}^T
    =\begin{pmatrix}
    0\\\textbf{\textit{f}}_g\\\textbf{\textit{v}}\cdot \textbf{\textit{f}}_g\\0
    \end{pmatrix},
\end{equation}
where, \textbf{\textit{f}}$_g$ is an acceleration term (e.g., due to gravity).

The \gPLUTO\, code is the GPU-accelerated C++ re-design \citep{Rossazza2025} of the widely used PLUTO legacy code \citep{Mignone2007} for astrophysical fluid dynamics by means of the 
%\sout{OpenACC programming model\footnote{\texttt{OpenACC:}  \url{https://www.openacc.org/}}.}
OpenACC\footnote{\texttt{OpenACC:}  \url{https://www.openacc.org/}} and OpenMP programming models.
%
%\color{red}
%[MR: CAN YOU PLEASE DOUBLE CHECK THIS SECTION ? ]
In this approach, loops are decorated with compiler directives that instruct the compiler on how to target GPU execution, leaving the low-level handling of threads, memory management, and scheduling to the compiler. 
This design enables the same codebase to be compiled and executed seamlessly on both CPUs (host) and GPUs (device).
In \gPLUTO, all the computation is executed directly by GPUs, while CPUs manage global control, memory allocation, and I/O operations. 
In fact, since GPU and CPU operate on their respective dedicated memory spaces in discrete architectures, data transfers are associated with significant overhead.
Therefore, to maximally utilize the parallel efficiency provided by GPUs, it is of utmost importance to ensure device data residency at all times, relegating host copies to the greatest extent.
In this context, loops and arrays have been designed to achieve coalesced access to memory, which means to ensure that multiple consecutive threads access adjacent memory spaces. 
On top, multi-GPU domain decomposition is implemented through MPI\footnote{\texttt{MPI:} \url{https://www.mpi-forum.org/}} with asynchronous communication, enabling efficient data exchange across neighboring subdomains. 

We note that the implementation of the relativistic MHD module in \gPLUTO\, adopts the same parallelization strategy as the non-relativistic version presented in \cite{Rossazza2025}.
Despite the moderately higher per-cell computational cost associated with primitive-variable recovery and flux evaluations, the increased arithmetic intensity leads to equally good (and in some cases improved) GPU performance.
%\color{black}

\subsection{Simulation setup for jet propagation}
\begin{figure*}
	\centering 
	\includegraphics[width=\textwidth]{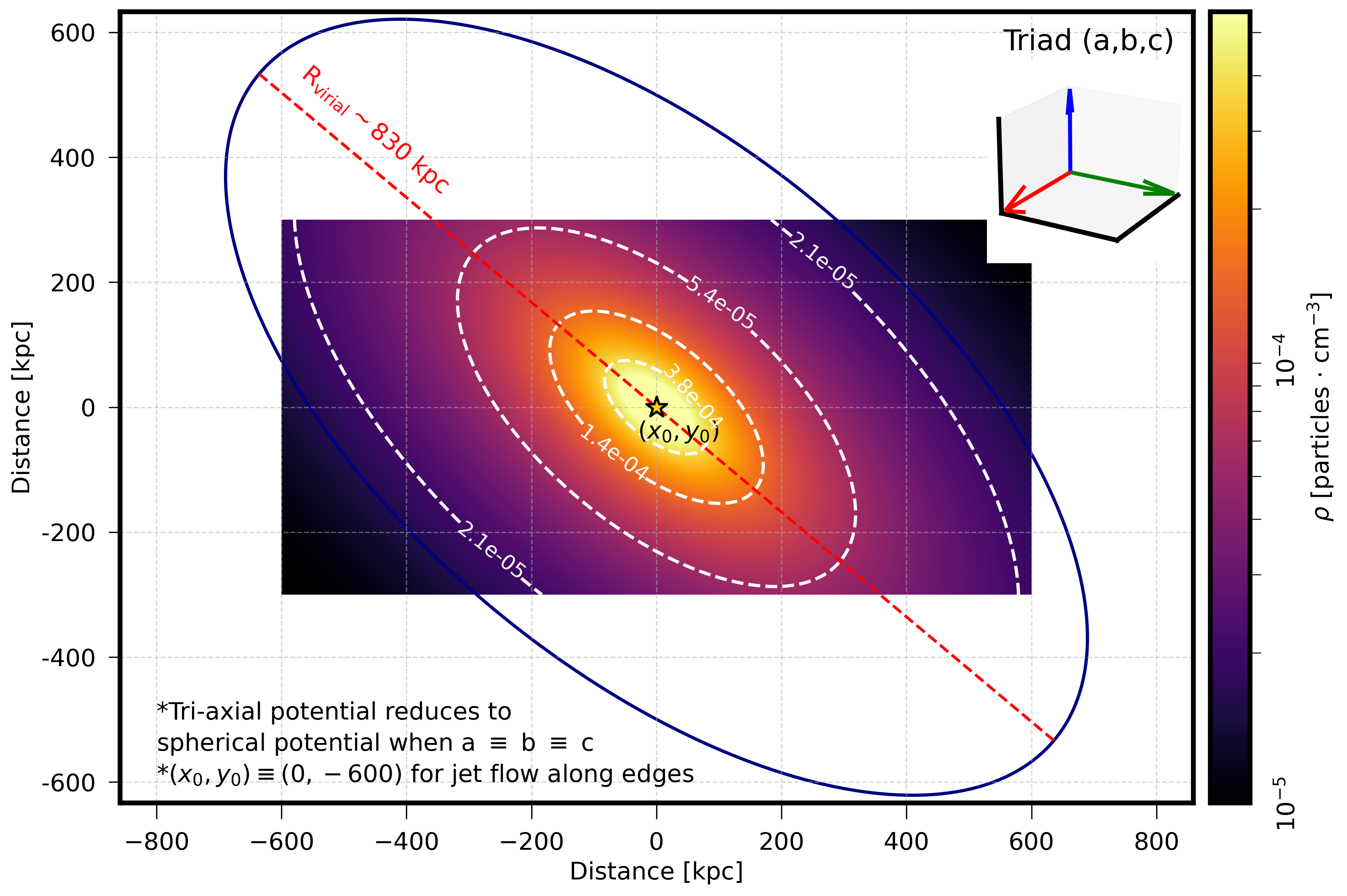}	
	\caption{Sliced representation of the triaxial density profile (mimicked through the triad) of a cosmic galaxy group (with core radii a, b, c, and virial radius $R_{\rm virial}$), representing the ambient medium through which extragalactic jets are injected and propagated in this study. Jets are primarily launched from the center (``GRG\_center''), while one case explores injection along the periphery (``GRG\_edge''), where the ambient medium is offset by $-\, 600$ kpc. The tilt of the medium relative to the jet propagation axis (along horizontal path) and its triaxiality demonstrate how we can implement symmetry-breaking in the environment to produce more realistic propagation scenarios (``GRG\_asymmetry''). White dashed contour lines, labeled inline, highlight the density stratification relevant to understanding jet-medium interactions.} 
	\label{fig:Initial_setup}
\end{figure*}

%%%%%%%%%%%%%%%%%%%%%%%%%%%%%%%%%%%%%%%%%%%%%%%%%%
In this study, we model the propagation of extragalactic jets within a poor galaxy group. 
This ambient medium is adopted following \citet{Giri2025_GRGI}, representing a triaxial galaxy group (see, Fig.~\ref{fig:Initial_setup}), parameterized through the core radii $a$, $b$, and $c$, with $a = b = c$ for most cases, corresponding to an effectively spherical medium, and $a \neq b = c$ for one case to investigate the effects of environmental asymmetry on jet propagation. 
The density distribution is described using a Cartesian form of the generalized King profile,  
\begin{equation}
\rho(x,y,z) = \frac{\rho_0}{\left[1 + \left(\frac{x-x_0}{a}\right)^2 + \left(\frac{y-y_0}{b}\right)^2 + \left(\frac{z}{c}\right)^2 \right]^{3\beta/2}},
\end{equation}  
where $\rho_0 = 0.001~\mathrm{amu\,cm^{-3}}$, $\beta = 0.55$ and the core radii $a, b, c$ range between 33 -- 66 kpc, producing a virial radius ($R_{\rm virial}$) of approximately 830 kpc, raising $R_{\rm 500} \approx R_{\rm virial}/2 \approx 415$ kpc \citep{Lovisari2015}. 
$R_{\rm 500}$ is defined as the radius within which the average density of the medium is 500 times the critical density of the Universe at the present epoch.
The ambient medium is formulated to be in static equilibrium, where the pressure gradient balances the gravitational force. 
The pressure ($P_g$) follows the adopted density profile, and the gravitational acceleration ($\mathbf{g}$) is obtained from the hydrostatic equilibrium condition: 
\begin{equation}
\nabla P_g = \rho\, \mathbf{g}.
\end{equation}
We note that the gravitational field is imposed as a fixed external potential and is not evolved self-consistently during the simulation evolution.
%{\color{red}[AM: I THINK THIS SHOULD BE $P_g$]}
The parameters $x_0$ and $y_0$ define the center of the ambient medium, which is $(0,0)$ for all cases. 
However, in one particular case (mentioned later), the center is shifted to $(x_0, y_0) = (0, -600~\mathrm{kpc})$ to allow the jet to propagate along the edges of the environment. 
%{\color{red}[AM: THE REFERENCE TO THE FIGURE IS MISLEADING, SINCE THE READER IMAGINES THAT THIS IS ONLY FOR THE SHIFTED VERSION...]}
In all simulations, the jet is injected from $(0,0,0)$ and flows around the $x$-axis throughout its evolution.

The jet is injected into the simulation box through a cylindrical nozzle  pointing along the positive $x$-axis, with initial Lorentz factor of $\gamma = 7$, representing a high-powered relativistic jet \citep[Fanaroff-Riley type - II;][]{Fanaroff1974}. 
The jet is light, with a density contrast of $10^{-5}$ relative to the ambient medium ($\rho_0$). 
In this study, we perform two categories of simulations. 

The first category consists of jets propagating up to 100~kpc, primarily to test the robustness of the \gPLUTO\, code in GPUs against CPU-based runs. 
These standard jet propagation extents in a large-scale cosmic environment are used for performance benchmarking. 
The jet is injected from $(0,0,0)$ through a cylinder of radius $1$~kpc, corresponding to a jet power of $5.6 \times 10^{44}$~erg\,s$^{-1}$.  
%{\color{red} [AM: SOUNDS LIKE A REPETITION TO ME. MAYBE WE CAN JOIN WITH THE PREVIOUS SENTENCE]}
We perform four resolution variants of this setup, corresponding to 1, 3, 8, and 15 grid cells per jet radius. 
These are labeled as follows, and the details of these runs are provided in Table~\ref{tab:Sim_setup}:
\begin{itemize}
    \item ``G100$^3$'': 1 cell per jet radius; 
    \item ``G300$^3$'': 3 cells per jet radius; 
    \item ``G800$^3$'': 8 cells per jet radius; 
    \item ``G1500$^3$'': 15 cells per jet radius.
\end{itemize}

\begin{table*}[htbp]
\makebox[\textwidth][c]{
\begin{tabular}{|l|c|c|c|c|c|}
\hline
\multicolumn{6}{|c|}{\textbf{Runs for performance benchmarking of gPLUTO}} \\
\hline
Simulation & Domain  & Grid & Resolution & Jet radius & Jet power \\
label & ([$X, \, Y, \, Z$] kpc$^3$) & cells & (cells / radius) & (kpc) & (erg $\cdot$ s$^{-1}$) \\
\hline
G100$^3$ & $[-10,\,100]$, $[-50,\,50]$, $[-50,\,50]$ & $110 \times 100 \times 100$ & 1 & 1.0 & $5.6 \times 10^{44}$ \\
G300$^3$ & $[-10,\,100]$, $[-50,\,50]$, $[-50,\,50]$ & $330 \times 300 \times 300$ & 3 & 1.0 & $5.6 \times 10^{44}$ \\
G800$^3$ & $[-10,\,100]$, $[-50,\,50]$, $[-50,\,50]$ & $880 \times 800 \times 800$ & 8 & 1.0 & $5.6 \times 10^{44}$ \\
G1500$^3$ & $[-10,\,100]$, $[-50,\,50]$, $[-50,\,50]$ & $1650 \times 1500 \times 1500$ & 15 & 1.0 & $5.6 \times 10^{44}$ \\
\hline
\multicolumn{6}{|c|}{\textbf{Runs for scientific applicability of gPLUTO}} \\
\hline
Simulation & Domain & Grid & Resolution & Jet radius & Jet power \\
label & ([$X, \, Y, \, Z$] kpc$^3$) & cells & (cells / radius) & (kpc) & (erg $\cdot$ s$^{-1}$) \\
\hline
GRG\_center & $[-5,\,995]$, $[-160,\,160]$, $[-160,\,160]$ & $2000 \times 640 \times 640$ & 5 & 2.5 & $3.5 \times 10^{45}$ \\
GRG\_edge & $[-5,\,995]$, $[-160,\,160]$, $[-160,\,160]$ & $2000 \times 640 \times 640$ & 5 & 2.5 & $3.5 \times 10^{45}$ \\
GRG\_asymmetry & $[-600,\,600]$, $[-300,\,300]$,  $[-300,\,300]$ & $2400 \times 1200 \times 1200$ & 5 & 2.5 & $3.5 \times 10^{45}$ \\
\hline
\end{tabular}
}
\caption{A concise overview of the seven simulation categories performed with \gPLUTO. The first four were executed on both GPUs and CPUs to benchmark the \gPLUTO’s performance across different hardware configurations (jets extending up to 100 kpc). The remaining three runs were carried out to test the scientific applicability and production readiness of \gPLUTO, on GPUs (jets extending to $\gtrsim 600$ kpc in one-sided propagation). Each column is duly labeled.}
\label{tab:Sim_setup}
\end{table*}

The second category of simulations was performed on much larger spatial scales ($\gtrsim 600$~kpc in one-sided jet propagation) to demonstrate the scientific applicability of \gPLUTO\, in modeling giant radio galaxies—a regime where high-resolution, full three-dimensional simulations remain exceedingly scarce. 
As GRGs are mostly powered by highly energetic jets \citep{Dabhade2020_LoTSS,Andernach2021,Simonte2024}, we adopt a jet radius of 2.5~kpc resolved with five grid cells, corresponding to a jet power of $3.5 \times 10^{45}$~erg\,s$^{-1}$. 

Three representative cases were explored.
The $1^{\rm st}$ one, labeled ``GRG\_centre'', models a jet launched from the center of a spherically symmetric environment.
The $2^{\rm nd}$ one, ``GRG\_edge'', represents a jet injected near the periphery of the large-scale environment, achieved by shifting the medium’s center to $y_0 = -600$~kpc such that the jet, launched from $(0,0,0)$, effectively propagates along the group’s edge. 
The $3^{\rm rd}$ case, ``GRG\_asymmetry'', investigates jet propagation through a triaxial environment where the major axis is tilted by $40^\circ$ relative to the jet propagation direction (which is positive $x-$direction for all the cases considered), enabling the study of asymmetric cocoon development (refer to Fig~\ref{fig:Initial_setup}).
%{\color{red}[AM: AS A GENERAL COMMENT, I DO NOT SEE A CLEAR CONNECTION BETWEEN THE ENVIRONMENT AND LAUNCHING ENGINE...]}

The simulations are briefly described below and summarized in Table~\ref{tab:Sim_setup} for their parametric details:
\begin{itemize}
    \item ``GRG\_center'': jet flows from the center of the (spherical) galaxy group medium;
    \item ``GRG\_edge'': jet flows from the edge of the (spherical) galaxy group medium;
    \item ``GRG\_asymmetry'': jet flows from the center of the (ellipsoidal) galaxy group medium (major axis inclined by $40^{\circ}$ to the jet axis).
\end{itemize}

We note that in all of our simulations, the jet 
%\sout{is injected in a magnetized form, with the initial magnetic field configuration prescribed to be purely toroidal. The field structure follows the prescription of \citet{Rossi2017}, and is given by}
carries a purely toroidal magnetic field, given by, \citep{Rossi2017}: 
\begin{equation}\label{Eq:B-field}
    B_y = B_0\, r\, \sin\phi, \quad
    B_z = B_0\, r\, \cos\phi,
\end{equation}
where $r$ and $\phi$ denote the polar coordinates in the $y$--$z$ plane, and $B_0$ sets the overall normalization of the field strength, which is fixed at $15\,\mu\mathrm{G}$ in all cases. 
The choice of a substantial toroidal component is well motivated by modern assumptions \citep[see, e.g.,][]{Mignone2010}, where such configurations are commonly invoked to drive jet acceleration and collimation. 
This particular geometry (in Eq.~\ref{Eq:B-field}) ensures that the magnetic field strength smoothly rises from zero at the jet axis to its maximum at the beam edge, effectively mimicking realistic magnetized jet launching conditions \citep[see, e.g.,][]{Lalakos2024}.

To track the evolution and spatial extent of the jet material, we employ a passive scalar (tracer), which is evolved according to a simple advection equation with the jet. 
The tracer values range from $1$ (indicating a cell fully occupied by jet plasma) to $0$ (indicating the absence of jet material), enabling a clear distinction between jet and ambient medium throughout the simulation.

\subsection{Computing infrastructure for the simulations}
All of our simulations were performed on the Leonardo\footnote{\texttt{CINECA -- Leonardo:} \url{https://www.hpc.cineca.it/systems/hardware/leonardo/}} supercomputing system hosted and managed by CINECA \citep{Turisini2023}. Leonardo consists of two main partitions\footnote{\texttt{System Architecture:} \url{https://docs.hpc.cineca.it/hpc/leonardo.html}}: the Booster partition, equipped with 4 GPUs and 32 CPU cores per node; and the Data-centric General Purpose (DCGP) partition, which provides CPU-only nodes with 112 CPU cores per node. A concise summary of node- and core-level architectural specifications is provided in Table~\ref{tab:HPC_leonardo}.
For benchmarking \gPLUTO, 
%\sout{we performed same simulation setups (Table~\ref{tab:Sim_setup}) across these configurations to assess performance scaling.} 
we employed the same setup (jet evolution followed up to 100 kpc) with different grid resolutions (see Table~\ref{tab:Sim_setup}) to assess performance scaling.
Specifically, the benchmark runs were first carried out on the Booster partition using GPUs, followed by equivalent CPU-only runs on the same partition, and finally repeated on the DCGP partition to compare against a higher-end CPU architecture (Table~\ref{tab:HPC_leonardo}). 
This enabled us to quantify the speedup of GPU acceleration relative to both equivalent and superior CPU resources. The scientific application runs (GRGs; Table~\ref{tab:Sim_setup}) were performed on the GPU-based Booster partition of the Leonardo.

\begin{table*}[htbp]
\begin{tabular}{l|c|c}
\hline
\textbf{Partition Name}          & DCGP              & Booster                   \\ \hline
\textbf{No. of nodes}            & 1536              & 3456                      \\
\textbf{CPU model}                                     & (Intel Sapphire Rapids) Intel Xeon Platinum 8480+ & (Intel Ice Lake) Intel Xeon Platinum 8358 \\
\textbf{CPU cores per node} & 112                                          & 32                  \\
\textbf{CPU Clock Rate}          & 2.6 GHz              &  2.0 GHz                          \\
\textbf{Memory}                  & 512(8x64) GB DDR5 & 512 GB DDR4               \\
%\textbf{Network}                 & 1x NDR200         & 2xNDR200                  \\
%\textbf{Node Bandwidth}          & 200 Gb/s           & 400 Gb/s                   \\
\textbf{Compiler / version}                            & g++ (GCC) 8.5.0 20210514 (Red Hat 8.5.0-18)                & mpicxx / nvc++ 24.5-1    \\
\textbf{No. of GPUs per node}    & –                 & 4                         \\
\textbf{GPU Model}               & –                 & NVIDIA Ampere100 custom\\
\textbf{GPU Memory}              & –                 & (HBM2e) 64 GB               \\
%\textbf{HBM Bandwidth}           & –                 & 2.0 TB/s                  \\
%\textbf{FP64 (Double Precision)} & –                 & 9.7 TFLOPS                \\ 
\hline
\end{tabular}
\caption{Key architectural specifications of the Leonardo supercomputing system (hosted and managed by CINECA), comparing the CPU-based DCGP partition and the GPU-enabled Booster partition. The listed specifications are relevant for interpreting the simulation performance comparisons presented in this work (Table~\ref{tab:speed_comparison}).}
\label{tab:HPC_leonardo}
\end{table*}

\section{Results}\label{Sec:Results}
%%%%%%%%%%%%%%%%%%%%%%%%%%%%%%%%%%%%%

We now present a performance benchmark between traditional CPU-based simulations and their GPU-accelerated counterparts (Section~\ref{Subsec:Performance Benchmarking of (CPU vs GPU) PLUTO}), providing a quantitative assessment of runtime gains achieved through GPU acceleration. 
Section~\ref{Subsec:Convergence analysis through resolution studies} presents a systematic optimization and convergence analysis between different simulations performed with GPU and CPU configurations. 
Furthermore, we demonstrate the application of \gPLUTO\, to model astrophysical scenarios where jet propagation physics remains relatively under-explored (Section~\ref{Subsec:Modeling Giant Radio Galaxies with GPU-Boosted Simulations}), highlighting its potential to enable more detailed and physically realistic studies.

\subsection{Performance benchmarking of (CPU vs GPU) \gPLUTO} 
\label{Subsec:Performance Benchmarking of (CPU vs GPU) PLUTO}
%%%%%%%%%%%%%%%%%%%%%%%%%%%%%%%%%%%%%%%%%%%%%%%%%%%%%%%%%%%%%%
We present here a detailed runtime comparison of individual jet propagation simulations performed using different versions of the \gPLUTO\, code, all executed on the Leonardo HPC system (see Table~\ref{tab:HPC_leonardo}).

We begin with the lowest resolution setup, $G100^3$, which is intentionally designed as a lightweight validation and entry-level benchmark rather than a production-scale performance test. This configuration is meant to assess code correctness and basic performance trends using minimal computational resources, representative of what a new user might access on a local workstation or small test allocation before scaling to full HPC runs. When executed on a single GPU device, this test completed in approximately 2 minutes, whereas the same problem required more than 170 minutes when run on a single CPU core. While this comparison is not intended to represent a realistic allocation-unit benchmark, it clearly demonstrates the intrinsic computational advantage of GPU offloading for this class of problems.

%which clearly illustrates the performance advantage of GPU acceleration already on a single core. When executed on a single GPU, this simulation completed in approximately 2 minutes, compared to more than 170 minutes for the equivalent CPU-only configurations. 
%This dramatic, two-order-of-magnitude reduction in runtime already highlights the transformative potential of GPU acceleration in enabling richer and more dynamic modeling of astrophysical jets.

Next, we consider the $G300^3$ configuration, which we adopt as our baseline resolution.
The performance difference between GPU- and CPU-based runs remains substantial. 
When comparing a single GPU-enabled node (with 4 GPUs) to a full CPU node with 32 cores (partition: Booster), the GPU run exhibits a significant speed-up, reducing the runtime by more than an order of magnitude. 
Even when benchmarked against a high-end state-of-the-art CPU node with 112 cores on the same HPC (partition: DCGP), the GPU-accelerated run still outperforms it by a factor of $\sim 14$. 
When compared to the 32-core CPU node configuration, the speed-up factor increases to approximately $\sim 32$. 
These results robustly demonstrate that GPU acceleration is not merely competitive but decisively advantageous for time-critical, high-resolution jet simulations.

A comparison of the final jet morphologies (2D density slices) at 6.52 Myr for ``$G300^3$'' runs in different resource settings is presented in the inset of the Table~\ref{tab:speed_comparison}, providing a visual reference to assess the consistency between configurations. 
The close resemblance between the outputs of GPU-accelerated and CPU-only simulations validates the \gPLUTO\, implementation.

The remarkable performance gains observed in the lower- and baseline-resolution runs motivated us to further test the capabilities of the code at more conventional resolutions—specifically, the ``$G800^3$'' case, corresponding to eight grid cells per jet radius. 
This configuration is computationally demanding and typically enters into a challenging regime for traditional CPU-based simulations. 
As shown in Table~\ref{tab:speed_comparison}, the GPU-accelerated run completed in approximately 5 hours, whereas even with 2.7 times longer wall-clock time, CPU-only runs with 32 cores per node and 112 cores per node were unable to reach the same marker point (i.e., 6.52 Myr). In terms of jet evolution time, the CPU-based runs lagged behind by 3.92 Myr and 1.63 Myr, respectively.

When comparing the achieved marker points of these simulations (Table~\ref{tab:speed_comparison}), the GPU-enabled run shows roughly a factor of $4$ speedup over the high-core (112) CPU-node configuration (at the third data saving interval, i.e., when the simulations reach 4.89 Myr) and nearly a factor of $\sim 7$ speedup relative to the 32-core CPU-nodes (at the first data-saving interval i.e., at 1.63 Myr)—a performance gap that is expected to widen as the simulation progresses (as evaluated from other grid distribution cases; Table~\ref{tab:speed_comparison}).
To contextualize the efficiency of the GPU-based runs and its production readiness, we also include an inset image (2D slice of density) showing the morphological and spatial resemblance between the ``$G800^3$'' and ``$G300^3$'' \gPLUTO\, outputs, illustrating that the accelerated performance does not compromise the physical fidelity of the simulations.

Building on these encouraging results, we extended our investigation to assess the performance of \gPLUTO\, for high-resolution simulations with 15 grid cells per jet radius (``$G1500^3$'')—a resolution that meets contemporary requirements for detailed dynamical and emission modeling in jet physics. 
Consistently with the lower-resolution cases, \gPLUTO\, in GPU-enabled nodes continues to significantly outperform even the high-core-count CPU only runs when the same number of nodes is used (refer to Table~\ref{tab:speed_comparison}), demonstrating its robustness and scalability for computationally demanding astrophysical simulations.

In summary, our performance benchmarking clearly demonstrates that GPU acceleration delivers a transformative improvement in simulation efficiency for jet propagation studies. 
Across all tested resolutions—from coarse (``$G100^3$'') to high-resolution (``$G1500^3$'') setups— \gPLUTO\, runs in GPU-enabled nodes consistently outperforms full state-of-the-art CPU-only configurations by factors ranging from approximately an order of magnitude to more than a factor of $\sim 30$.
These speedups are achieved without compromising the physical accuracy or morphological fidelity of the simulated jets, as confirmed by direct visual and quantitative comparisons. 
This substantial gain in computational efficiency enables higher-resolution, longer-duration, and more physically sophisticated simulations within feasible wall-clock times, establishing GPU acceleration as a powerful and scalable tool for advancing theoretical modeling of astrophysical jet phenomena.

\begin{table*}
\centering
\renewcommand{\arraystretch}{1.3} % spacing between rows
\begin{tabular}{|l|c|c|c|c!{\vrule width 2pt}c!{\vrule width 2pt}}
\hline
Leonardo -- CINECA & Sim. time & [G: $1500^3$] [20 N] & [G: $800^3$] [12 N] & [G: $300^3$] [1 N] & [G: $100^3$] [1 \texttt{GPU}]\\
(Resource specification) & (Myr) & Run time (min) & Run time (min) & Run time (min) & Run time (min)\\ \hline \hline

\multirow{4}{*}{\shortstack[l]{\texttt{GPU} partition (Booster) \\ $\left[\text{4 \texttt{GPUs}}\right]/\text{Node}$}} & 1.63  & 274.7 & 69.8 & 4.48 & 0.4\\ \cline{2-6}
 & 3.26  & 595.9 & 127.2 & 9.75 & 0.8 \\ \cline{2-6}
 & 4.89  & -- & 214.6 & 17.4 & 1.3 \\ \cline{2-6}
 & 6.52  & -- & {\setlength{\fboxrule}{1.2pt}\fcolorbox{OliveGreen}{white}{\makebox[2cm][c]{\strut 311.2}}} & {\setlength{\fboxrule}{1.2pt}\fcolorbox{OliveGreen}{white}{\makebox[2cm][c]{\strut 27.7}}} & 1.9 \\ \hline

\multicolumn{6}{|c|}{
\includegraphics[width=0.45\textwidth]{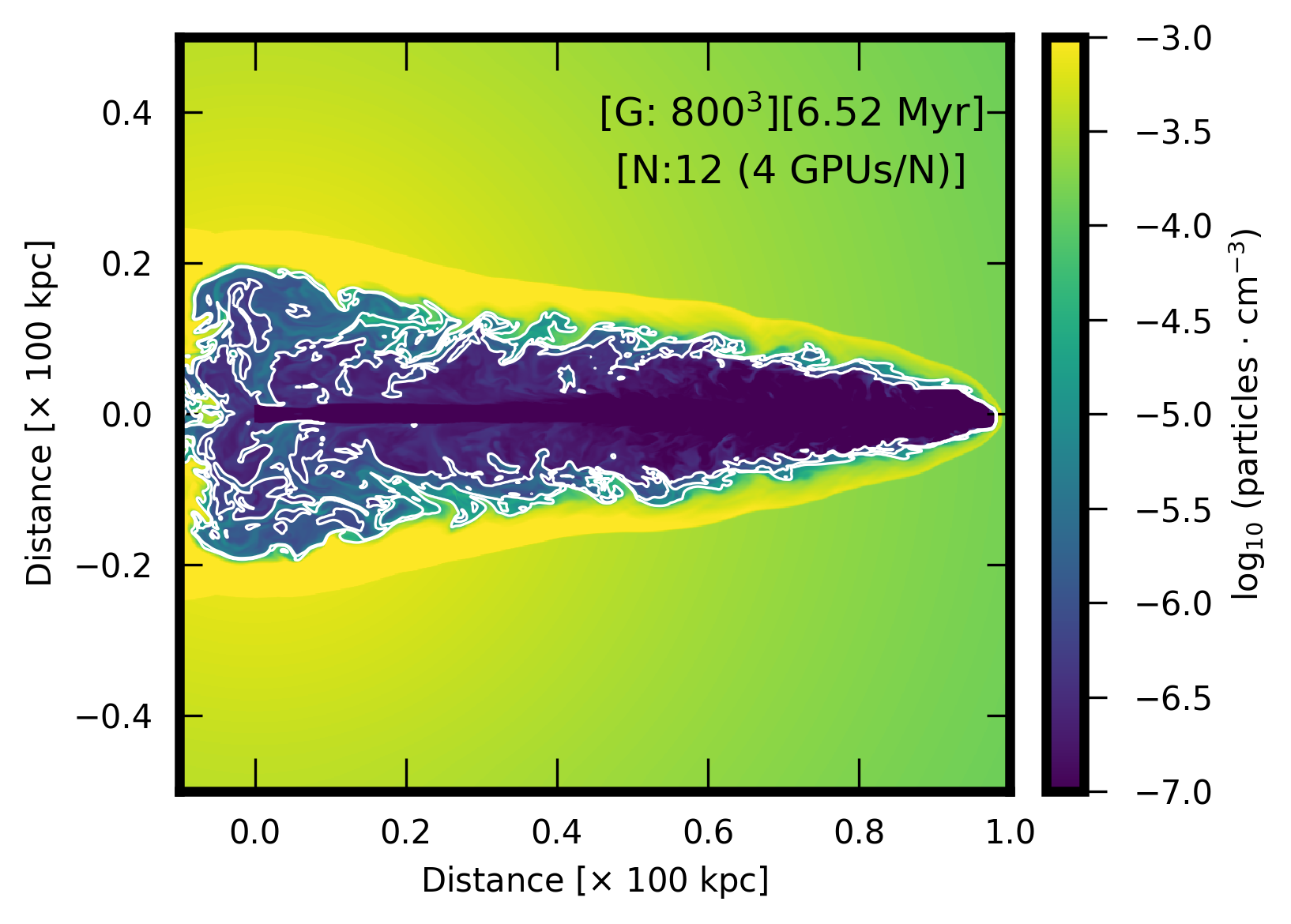} % replace with your figure
\includegraphics[width=0.45\textwidth]{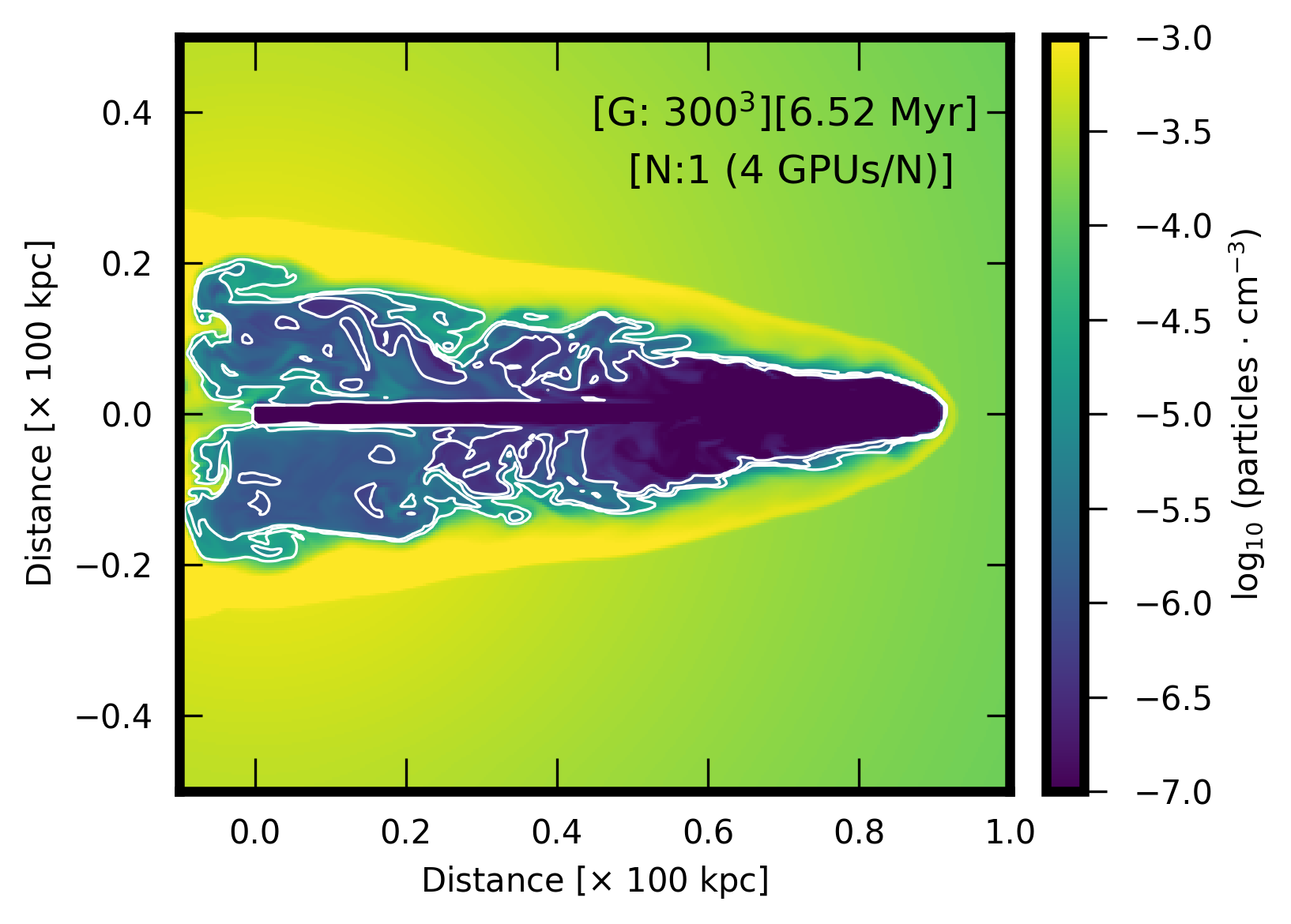} % replace with your figure
} \\ \hline

(Resource specification) & Sim. time & [G: $1500^3$] [20 N] & [G: $800^3$] [12 N] & [G: $300^3$] [1 N] & [G: $100^3$] [1 \texttt{CPU}]\\ \hline

\multirow{4}{*}{\shortstack[l]{\texttt{GPU} partition (Booster) \\ $\left[\text{32 \texttt{CPU} cores}\right]/\text{Node}$}} & 1.63 & -- & 474.1 & 110.8 & 30.2 \\ \cline{2-6}
 & 3.26 & -- & \parbox[c]{2.8cm}{\centering 888.4$^{*}$\\ \footnotesize{$^{*}$followed to 2.6 Myr}} & 290.2 & 71.2 \\ \cline{2-6}
 & 4.89 & -- & -- & 553.8 & 123.5 \\ \cline{2-6}
 & 6.52 & -- & -- & {\setlength{\fboxrule}{1.2pt}\fcolorbox{OliveGreen}{white}{\makebox[2cm][c]{\strut 894.7}}} & 188.4 \\ \hline

\multicolumn{6}{|c|}{
\includegraphics[width=0.45\textwidth]{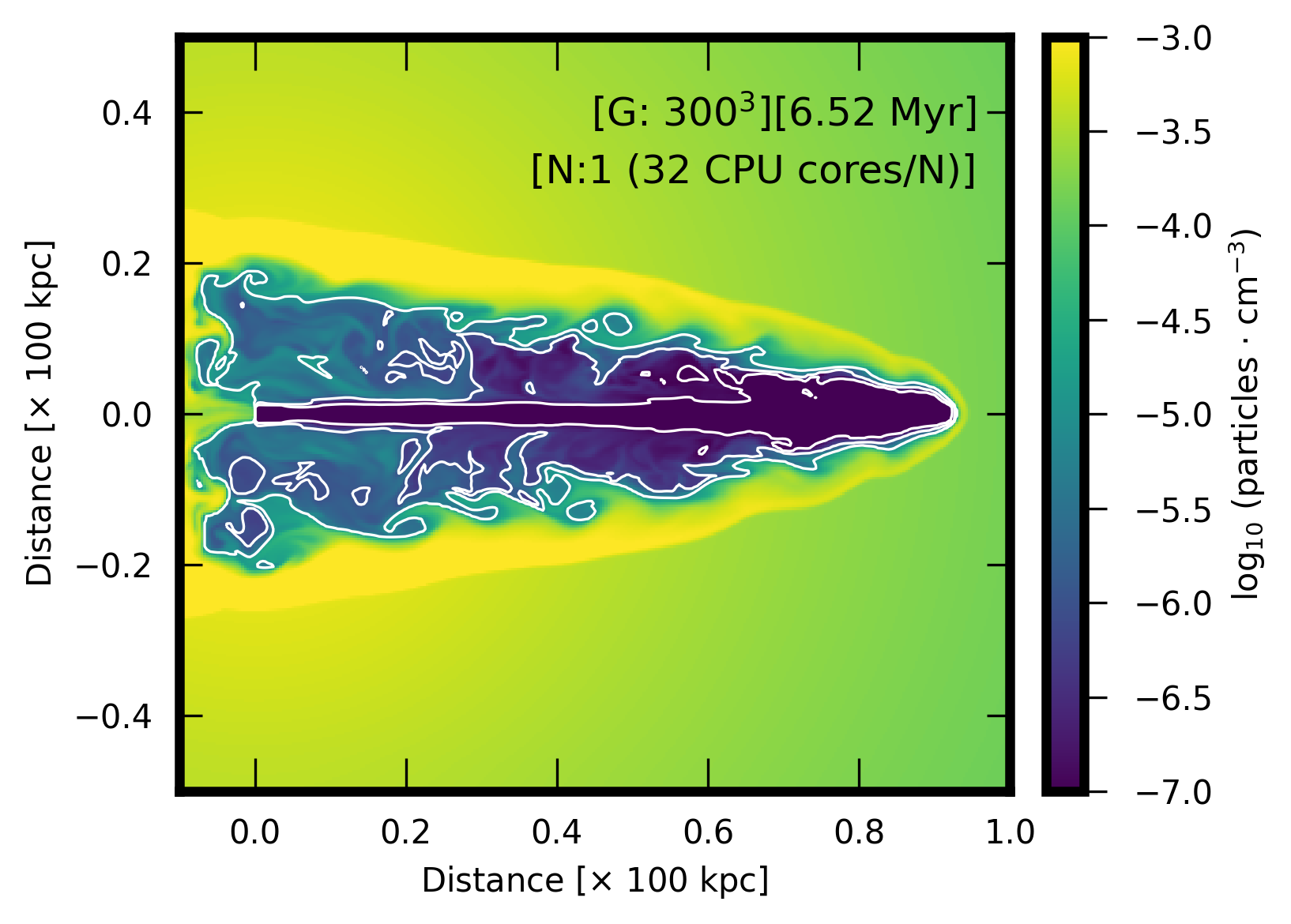}
\includegraphics[width=0.45\textwidth]{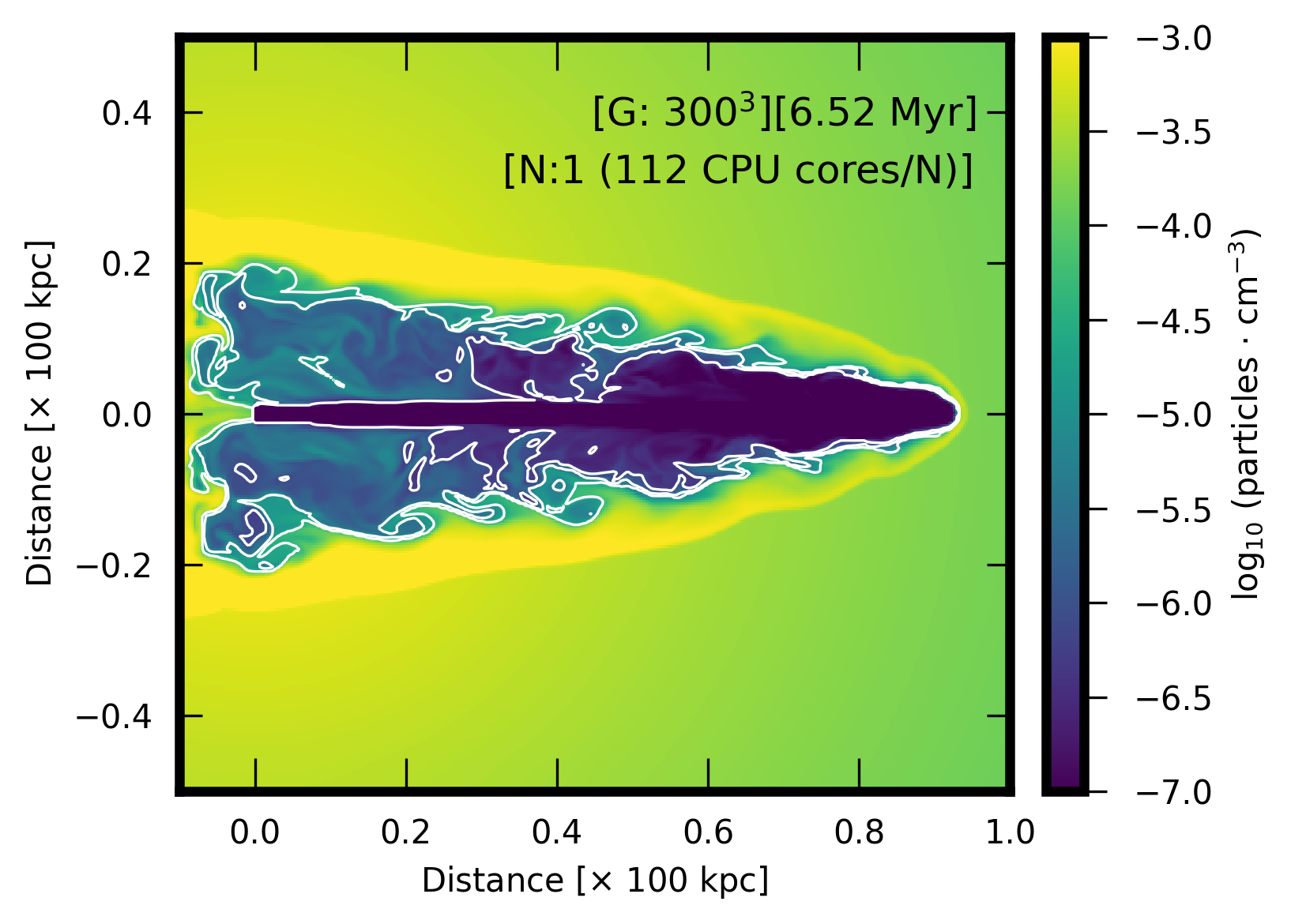}% replace with your figure
} \\ \hline

(Resource specification) & Sim. time & [G: $1500^3$] [20 N] & [G: $800^3$] [12 N] & [G: $300^3$] [1 N] & [G: $100^3$] [1 \texttt{CPU}] \\ \hline

\multirow{4}{*}{\shortstack[l]{\texttt{CPU} partition (DCGP) \\ $\left[\text{112 \texttt{CPU} cores}\right]/\text{Node}$}} & 1.63 & \parbox[c]{2.8cm}{\centering 856.4$^{*}$\\ \footnotesize{$^{*}$followed to 1.2 Myr}} & 158.9 & 47.1 & 31.0 \\ \cline{2-6}
 & 3.26 & -- & 446.4 & 123.7 & 73.1 \\ \cline{2-6}
 & 4.89 & -- & 856.1 & 233.0 & 126.4 \\ \cline{2-6}
 & 6.52 & -- & -- & {\setlength{\fboxrule}{1.2pt}\fcolorbox{OliveGreen}{white}{\makebox[2cm][c]{\strut 371.1}}} & 194.0 \\ \hline

%\multicolumn{4}{|c|}{
%\includegraphics[width=0.9\textwidth]{example-image} % replace with your figure
%} \\ 
\hline

\end{tabular}
\caption{Tabular summary of simulations performed with \gPLUTO\, code using different CPU and GPU configurations. \textit{Col. 1} lists the resource setups on the Leonardo HPC system. \textit{Col. 2} shows marker points (data-saving intervals) corresponding to jet evolution time in Myr. \textit{Col. 3–6} give runtimes (minute) to reach each marker for varying grid sizes (G) and node allocations (N). Representative 2D slices from selected runs (green boxes) enable visual comparison. Dashed and asterisks markers denote partially completed runs (followed till a conclusion can be made). Overall, the table highlights the performance advantage of GPU-accelerated PLUTO over (even high-end) CPU-only runs. The last column is visually separated to denote an entry-level validation test performed on a single GPU device versus on a single CPU core.}
\label{tab:speed_comparison}
\end{table*}

\subsection{Convergence analysis through resolution studies}
\label{Subsec:Convergence analysis through resolution studies}
%%%%%%%%%%%%%%%%%%%%%%%%%%%%%%%%%%%%%%%%%%%%%%%%%%%%%%%%%%%%%%%
It is particularly insightful to examine the similarities obtained for 100 kpc jet propagation across different CPU and GPU configurations, each performed at varying resolutions. 
This comparison not only provides a clearer perspective on the consistency between CPU and GPU results but also underscores the critical role of resolution in accurately modeling astrophysical jet dynamics.

\begin{figure*}
	\centering 
	\includegraphics[width=\columnwidth]{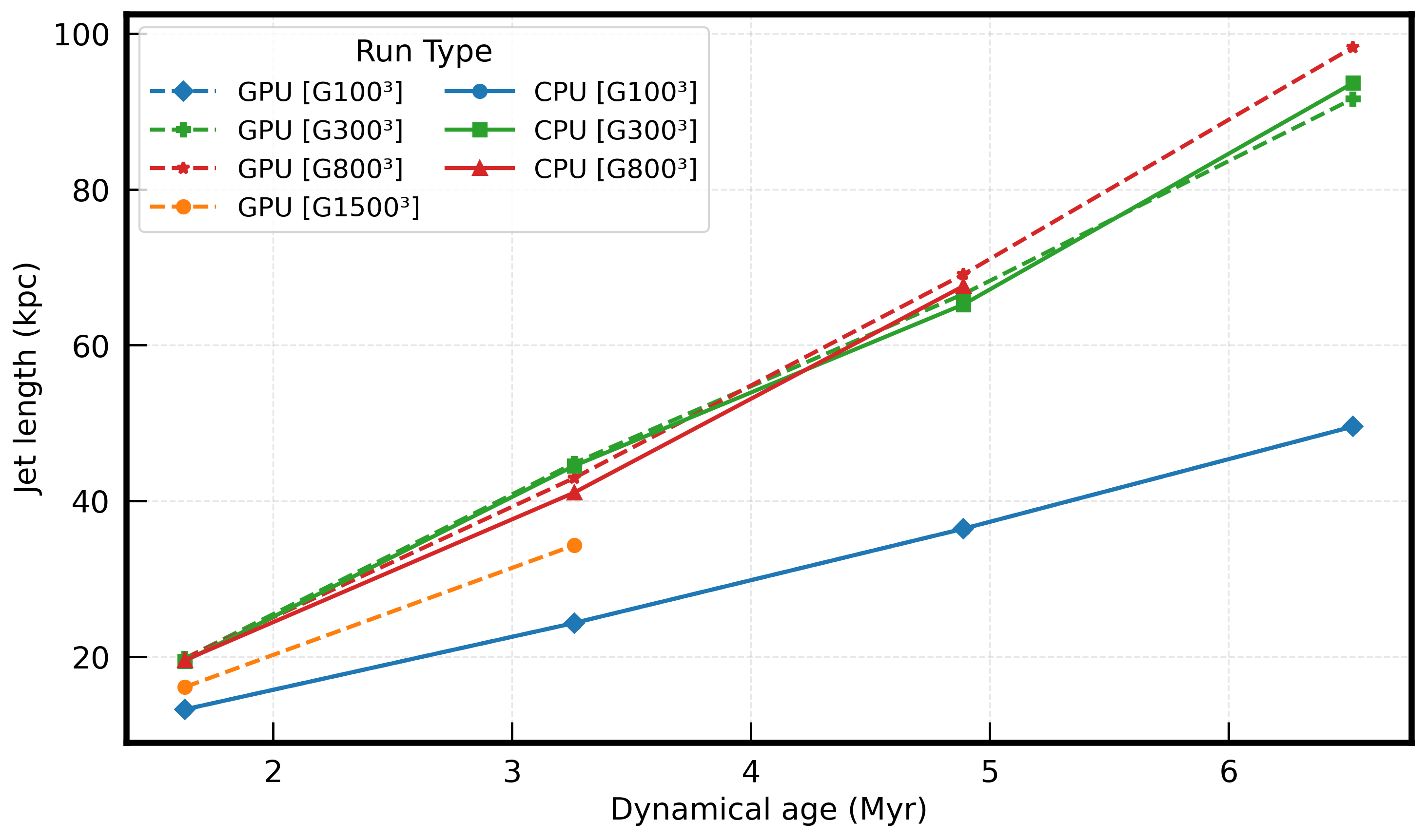}
    \includegraphics[width=\columnwidth]{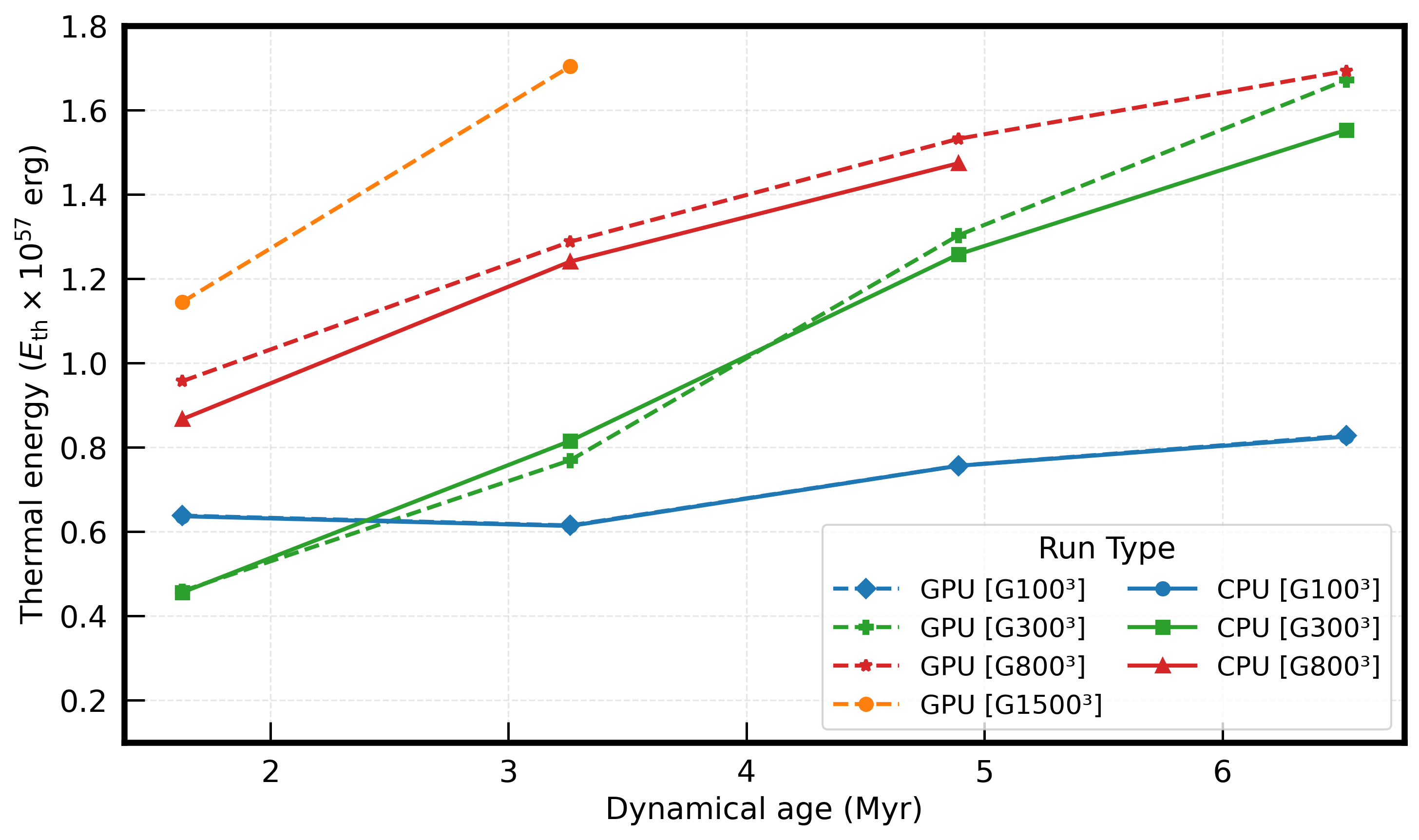}
	\caption{\footnotesize Comparison of \gPLUTO\, CPU-only and GPU simulations showing the jet length evolution as a function of dynamical age (\textit{left}) and the evolution of the thermal energy within the jet cocoon as a function of dynamical age (\textit{right}), highlighting the production readiness of \gPLUTO\, (in GPUs). The plot also illustrates the importance of resolution for accurately capturing jet propagation physics.} 
	\label{fig:Resolution_convergence}
\end{figure*}

Fig.~\ref{fig:Resolution_convergence} (\textit{left}) shows the jet propagation length as a function of dynamical age, measured by the maximum extent where the jet tracer exceeds $10^{-7}$, a standard proxy for the jet lobe size \citep{Mukherjee2020}. Fig.~\ref{fig:Resolution_convergence} (\textit{right}) shows the temporal evolution of the cocoon’s thermal energy, a genuinely volumetric diagnostic (region with jet tracer $\geq 10^{-7}$) that more faithfully traces the structural evolution of the system—such as global lobe expansion, internal adiabatic compressions, and particle mixing—than the approximate 1-D evolution of the jet length.
Comparing runs at the same resolution between CPU-only and GPU-enabled \gPLUTO, from $G100^3$ to $G800^3$, reveals a near-identical evolution trajectory. 

This demonstrates both the reliability and production readiness of the GPU-accelerated PLUTO code (\gPLUTO), while highlighting its significant wall-clock speed-up (Table~\ref{tab:speed_comparison}), which substantially enhances the efficiency of magneto-hydrodynamical simulations in jet astrophysics—a domain where PLUTO (legacy code) remained a widely adopted computational tool.

Resolution plays a critical role in accurately capturing jet propagation dynamics, as demonstrated in Fig.~\ref{fig:Resolution_convergence} (\textit{left}). 
In the lowest-resolution runs, the jet head exhibits a noticeable lag compared to higher-resolution counterparts, which could lead to misinterpretation of jet evolution. While runs with 3 and 8 cells per jet radius slightly overestimate propagation relative to the 15-cell reference, the differences in these cases are not extreme however. These discrepancies largely reflect how well the simulation resolves the pressure gradients of the ambient medium and the development of magneto-hydrodynamical instabilities within the jet beam and at the cocoon edges \citep{Giri2025_GRGI}. Fig.~\ref{fig:Resolution_convergence} (\textit{right}) presents a more informative diagnostic of the cocoon evolution. The lowest-resolution run fails to reproduce the evolutionary trends observed at higher resolutions, exhibiting only weak temporal variations. For intermediate resolutions (between 3 and 8 cells per jet radius), a modest discrepancy appears during the early stages, reflecting the different ability to resolve the initially compact jet structure between the cases; however, the thermal energy evolution converges at later times as the jet expands. The highest-resolution run (15 cells per jet radius) displays a consistent evolutionary behavior, characterized by a slower expansion rate (see Fig.~\ref{fig:Resolution_convergence} (\textit{left})) and correspondingly higher internal pressure within the cocoon.

This analysis underscores the importance of adequately resolving the jet radius and performing convergence tests to ensure reliable and physically meaningful results in jet dynamics simulations.

\subsection{Modeling giant radio galaxies with GPU-boosted simulations}
\label{Subsec:Modeling Giant Radio Galaxies with GPU-Boosted Simulations}
%%%%%%%%%%%%%%%%%%%%%%%%%%%%%%%%%%%%%%%%%%%%%%%%%%%%%%%%%%%%%%%%%%

\begin{figure*}
	\centering 
	\includegraphics[width=\textwidth]{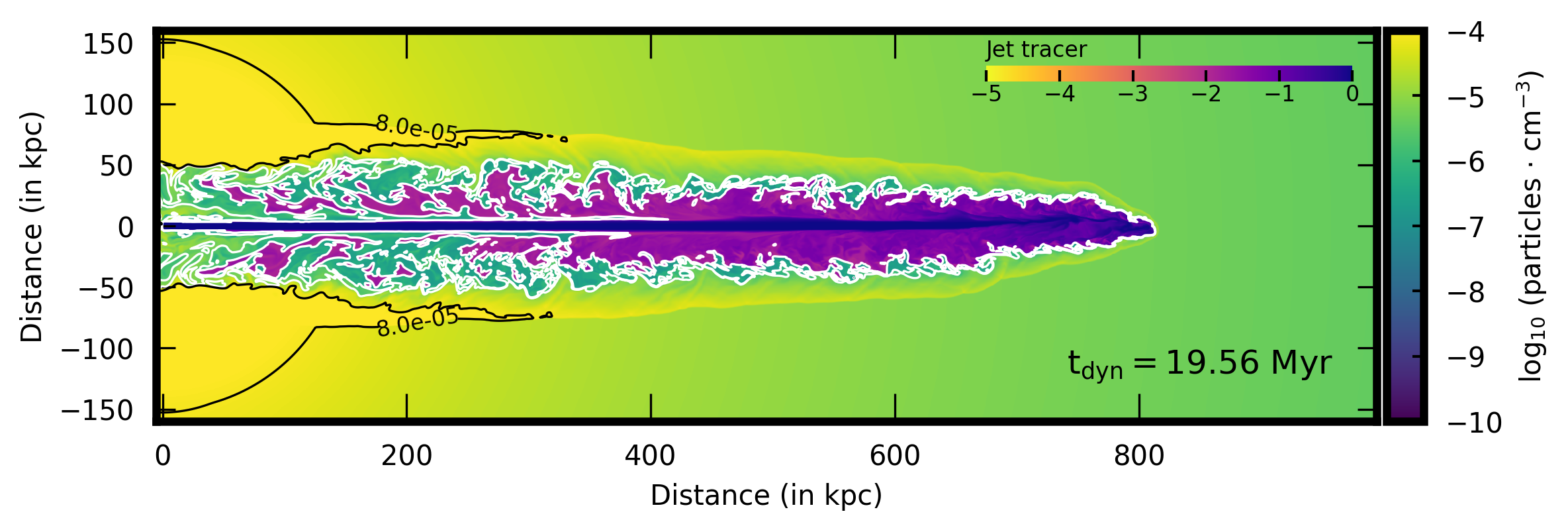}	
	\caption{\footnotesize Propagation of a jet reaching giant radio galaxy scales after originating from the center of the large-scale cosmic environment (represented here by a galaxy group, as also inferred by the black density contours). The structure of the jet cocoon is clearly captured using the jet tracer, displayed both as contours (white) and a colormap, and is overplotted on the overall density distribution (all presented in $x-y$ slice at $z = 0$) of the simulated system to illustrate the connection between the jet lobes and the surrounding medium. The evolution time of the simulation is also indicated as inset.} 
	\label{fig:GRG_st_centre}
\end{figure*}

\begin{figure*}
	\centering 
	\includegraphics[width=\textwidth]{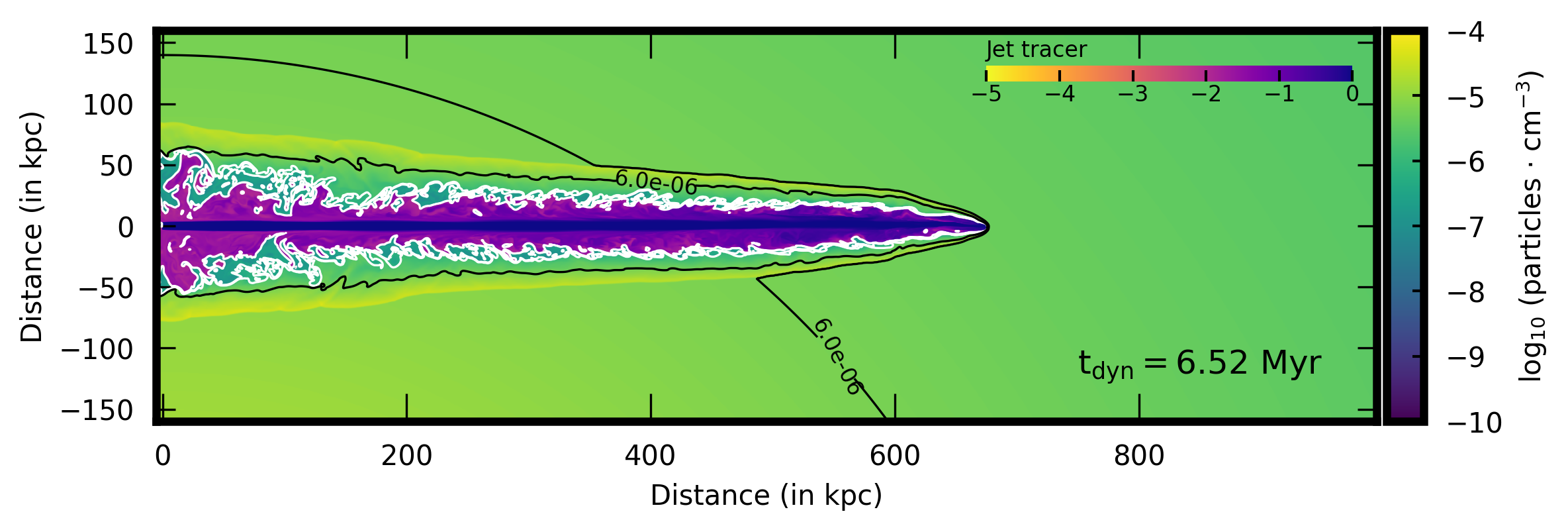}	
	\caption{\footnotesize Similar to Fig.~\ref{fig:GRG_st_centre}, but in this case the jet is injected from the outskirts of the large-scale cosmic environment, a configuration often associated with giant radio galaxies. The jet propagates rapidly, reaching approximately 700 kpc in just 6.52 Myr. This fast propagation is important for understanding recent observations of giant jets extending to 5 Mpc and beyond, providing insights into their formation and evolution.} 
	\label{fig:GRG_edge}
\end{figure*}

\begin{figure*}
	\centering 
	\includegraphics[width=\textwidth]{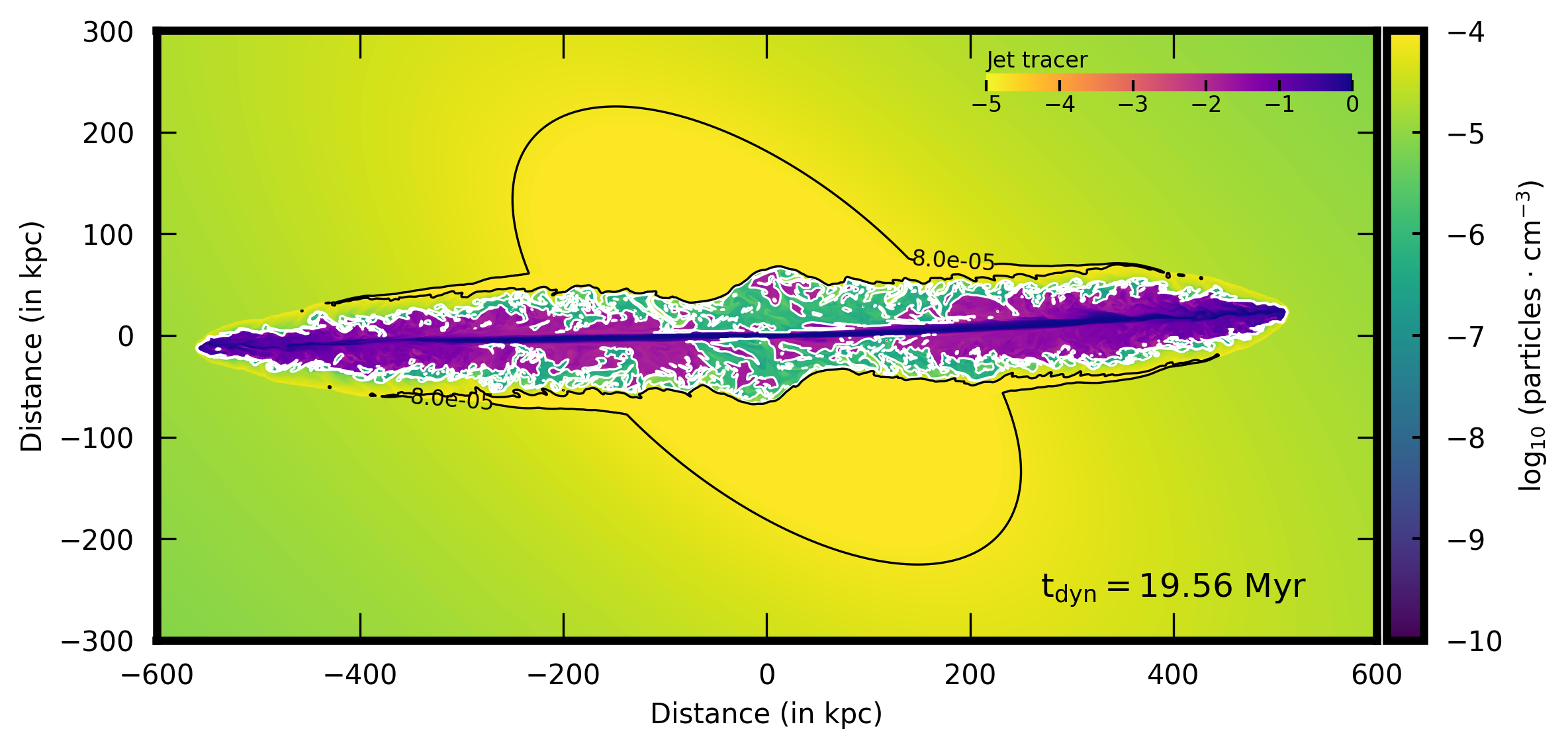}	
	\caption{\footnotesize Same as Fig.~\ref{fig:GRG_st_centre}, but in this case the jet is injected from the center of a triaxial medium to examine its propagation under asymmetric environmental conditions. Unlike the previous giant radio galaxy cases, this jet is bidirectional, allowing us to assess the impact of reduced symmetry on the evolution of the system. The propagation speed is slightly lower here due to the increased environmental resistance of the triaxial structure. The limb-like cocoon feature observed in the jet-base region arises from the backflowing plasma preferentially extending along the steepest pressure gradient. Such a feature is expected to become even more prominent in lower power jets, potentially giving rise to complex morphologies such as X-shaped radio galaxies.} 
	\label{fig:XRG_GRG}
\end{figure*}

Given the substantial reduction in computational time achieved with \gPLUTO, it is natural to extend its application to modeling giant radio galaxies—an area where high-resolution simulations remain limited. 
We have therefore performed simulations at 5 cells per jet radius resolutions for representative giant radio galaxy cases (Table~\ref{tab:Sim_setup}), as illustrated in Fig.~\ref{fig:GRG_st_centre}, \ref{fig:GRG_edge}, and \ref{fig:XRG_GRG}, to demonstrate the practical applicability and efficiency of GPU-accelerated modeling in this domain.

Fig.~\ref{fig:GRG_st_centre} shows the propagation of a powerful jet launched from the center of a galaxy group (``GRG\_center'' case). 
As it encounters strong environmental resistance, the jet reaches a one-sided distance of $\sim 800$ kpc in about 19.6 Myr, demonstrating how GRGs with total extents of $\sim 1.5$ Mpc can originate from central host galaxies, often the brightest members of large-scale cosmic structures \citep{Sankhyayan2024}. 
The sustained collimation of the jet at 800 kpc (Fig.~\ref{fig:GRG_st_centre}) indicates its potential for further growth, offering insight into how GRGs larger than 3 Mpc can also emerge from central environments, shedding light on a contemporary finding \citep{Andernach2025}.

For the case shown in Fig.~\ref{fig:GRG_edge}, where the jet is launched from the outskirts of the same galaxy group ($\sim 600$ kpc from the center), the lower environmental resistance allows the jet to reach a distance of $\sim 700$ kpc in just 6.52 Myr (``GRG\_edge'' case). 
Such rapidly propagating jets, if evolved over longer timescales, could likely explain GRGs with remarkable extents up to 5 Mpc or more (Giri et al. \textit{in prep.}). 
These sources exhibit well-collimated beams enclosed within narrow cocoons—features characteristic of observed GRGs \citep{Machalski2008,Oei2024_7Mpc}, relevant to what we have obtained in Fig.~\ref{fig:GRG_edge}. 
The short propagation time obtained in this case \citep[also, see,][]{Jamrozy2008} supports the growing evidence that many GRGs originate or expand along the peripheries of dense cosmic environments \citep{Subrahmanyan2008}, often evolving in filaments or voids \citep{Oei2024_fiaments,Mahato2025}. 
This scenario aligns with recent observational trends showing a rapid increase in GRG detections from modern surveys. 
The inferred sky density of these systems is becoming comparable to that of luminous non-giant radio galaxies \citep{Mostert2024}, indicating that GRGs are not rare or exceptional objects but may in fact be a common population in the radio sky.

To explore GRG evolution in a slightly asymmetric environment, Fig.~\ref{fig:XRG_GRG} shows a jet with similar power and density profile as in the previous cases, propagating through a mildly triaxial medium (``GRG\_asymmetry'' case). 
The jet reaches $\sim 600$ kpc in 19.6 Myr, slowed by increased ambient resistance along and around the major axis of the triaxial density distribution. 
While the overall propagation dynamics resemble Fig.~\ref{fig:GRG_st_centre}, the head advance speed is slightly reduced. Notably, the backflowing plasma \citep[after originating at the jet head due to entropy imbalance, matter backflows;][]{Cielo2017} at the jet base shows a mild lateral extension along the minor axis due to pressure gradients. 
In lower-power jets, this backflow could extend further, potentially forming X-shaped morphologies \citep{Capetti2002,Rossi2017}. 
These dynamics are relevant for understanding the formation of GRG-XRG systems, which, though rare, are increasingly being observed and characterized in detail \citep{Saripalli2009,Cotton2020,Bruni2021}.

\section{Summary}\label{Sec:Summary}
In this work, we have presented an investigation of astrophysical jet propagation using the \gPLUTO\, code, comparing traditional CPU-based simulations with its recently developed GPU-accelerated version. 
Our performance benchmarks demonstrate that \gPLUTO\, consistently outperforms CPU-only runs across multiple resolutions—from coarse ($G100^3$) to high-resolution ($G1500^3$) setups—achieving speedups ranging from an order of magnitude to approximately over 30 without compromising physical fidelity. 
These results establish \gPLUTO\, as a production-ready tool for large-scale magneto-hydrodynamical simulations, enabling longer-duration, higher-resolution studies of jet evolution within feasible wall-clock times.

We further analyze convergence of jet simulations through resolution studies, highlighting the critical role of adequate spatial resolution in accurately capturing jet propagation dynamics. 
Low-resolution runs underestimate key jet dynamics, such as the jet-head advance speed and the evolution of internal energy, potentially misrepresenting the jet evolution, whereas higher-resolution runs converge toward consistent dynamical trajectories.

Following these results, leveraging the computational efficiency of \gPLUTO, we model representative giant radio galaxies under various environmental conditions—including central, peripheral, and triaxial host media. These simulations reproduce observed jet morphologies, propagation distances, and cocoon structures, providing insights into extreme GRG formation, collimation maintenance over Mpc scales, and the potential origins of GRG cocoon asymmetries. 

Overall, our study underscores the transformative potential of GPU-accelerated simulations, with \gPLUTO\, emerging as a powerful and scalable tool for modeling relativistic, magnetized astrophysical jets on modern exascale architectures.

\section*{Acknowledgements}
We acknowledge Istituto Nazionale di AstroFisica (INAF) for awarding this project access and computational time to the LEONARDO supercomputer, owned by the EuroHPC Joint Undertaking, hosted by CINECA (Italy) and the LEONARDO consortium. This paper is supported by the Fondazione ICSC, Spoke 3 Astrophysics and Cosmos Observations and National Recovery and Resilience Plan (Piano Nazionale di Ripresa e Resilienza, PNRR), Project ID CN 00000013 “Italian Research Center on High-Performance Computing, Big Data and Quantum Computing” funded by MUR Missione 4 Componente 2 Investimento 1.4: Potenziamento strutture di ricerca e creazione di “campioni nazionali di R\&S (M4C2-19)” - Next Generation EU (NGEU). This work has also received funding from the European High Performance Computing Joint Undertaking (JU) and Belgium, Czech Republic, France, Germany, Greece, Italy, Norway, and Spain under grant agreement No 101093441 (SPACE).

%\appendix
%\section{Appendix title 1}

\bibliographystyle{elsarticle-harv} 
\bibliography{example}



%% For example:

%% \bibitem[Aladro et al.(2015)]{Aladro15} Aladro, R., Martín, S., Riquelme, D., et al. 2015, \aas, 579, A101

%%\end{thebibliography}

\end{document}